\documentclass{article}
\begin{document}
\title{Cosmic Radio Jets}

\author{Paul J.\ Wiita\\
Department of Astrophysical Sciences, Princeton University,\\
Princeton NJ 08544-1001, USA; on leave from the\\
Department of
Physics and Astronomy, Georgia State University\\
wiita@chara.gsu.edu}
\maketitle
\begin{abstract}
Extragalactic radio sources, including quasars, are now typically 
understood as being produced by a pair of nearly symmetric, oppositely
directed relativistic jets.  While some these sources span megaparsecs, 
and are thus the largest physically connected structures in the
universe, emitting regions identified as jets have now been found on all scales down
to fractions of a parsec, and  jets appear to be a
common element of most (maybe all) types of active galactic nuclei
(AGN).  We first 
summarize key observations of different
classes of cosmic radio jets, and describe how they may be
connected. Theoretical models for the launching
and propagation of extragalactic jets are briefly described.  All of these
models assume a magnetized plasma, which typically amounts to only
a small fraction
of the accreted gas, is ejected from the vicinity of a supermassive
black hole. The extreme
complexity of the relevant physics has demanded numerical simulations
to examine non-linear effects on the stability of propagating jets,
and some recent results from these efforts are summarized.

\end{abstract}

\section{Introduction:  ``Ancient'' History}
The discovery of extragalactic jets must be considered one of the
triumphs of astrophysics, as it involves one of the relatively
few actual {\it predictions} in this fundamentally observation driven
science.  Early radio telescopes revealed the existence of
extragalactic
double radio sources such as Cygnus A (Jennison \& Das Gupta 1953),
and led to the discovery of quasars in the 1960's.  The combination
of non-thermal, essentially power-law, radio spectra, and often substantial
polarization, very quickly led to the emission mechanism
being identified as synchrotron radiation from relativistic
charged particles in helical orbits around magnetic field lines.
In the early 1970's
a few theoretical papers (Rees 1971, Longair et al.\ 1973,
Scheuer 1974, Blandford \& Rees
1974) proposed that these extremely large (typical scales of 100
kpc) and powerful (typical $L_R \sim 10^{44}~{\rm erg~s}^{-1}$) sources were fed
by jets emerging from the centers of elliptical galaxies.  Only with
the dynamic range provided by the Very Large Array and other radio
telescopes coming on line later in the 1970's were many of these jets actually
detected.  Competing models, involving blobs
(or ``plasmoids'' or ``plasmons'') of radio emitting plasma (e.g., Pacholczyk 1977)
or gravitational-slingshots that eject independent engines and
sources
of plasma (e.g., Saslaw, Valtonen \& Aarseth 1974) 
are now rarely considered.  This is because of 
the convincing evidence of nearly co-linear emission extending from less than
a pc to hundreds of kpc in many sources.  While
we shall not discuss these alternative scenarios any further, it is still
worthwhile to note that many of the observations of radio galaxies can
be accounted for by the slingshot model (Valtonen \& Hein\"am\"aki 1999).

The vast majority of these extragalactic radio sources can be
easily classified into one of two categories, as first noted by
Fanaroff \& Riley (1974), who carefully examined  the first set of synthesis maps. 
The weaker ones, or FR I's, with radio fluxes  $P_{{\rm 178} {\rm MHz}} <
2 \times 10^{25}h_{50}^{-2}$ W Hz$^{-1}$ sr$^{-1}$ (where $h_{50}$
is the Hubble constant in units of 50 km s$^{-1}$ Mpc$^{-1}$),  tend 
to have their brightest regions within the inner half of the
total extent of the radio source; i.e., much of the flux emerges
from the jets themselves, and the fluxes emanating from the two
jets are almost always comparable.  The FR I sources can be further
sub-classified into fairly symmetrical 
twin-jets, 
Wide Angle Tail, and Narrow Angle Tail (or head--tail) sources (e.g.,
Bridle \& Perley 1984). 
Many of the details of these diverse structures can be understood in terms
of interactions of the jets with the motions of, or irregularities
in, the density
and magnetic fields of the external media (interstellar, intracluster
and intergalactic) through which they propagate. 
Other FR I morphologies, particularly S-type symmetries,
 could be explained by the precession of
the jets axes or ejection of the jets from one component of a
binary AGN. 

More powerful sources, or FR II's, emit the bulk of their radio photons from
the outer portions of their structures.  Almost all of these are the
classical double sources, with most emission coming from lobes which
are usually well separated and extend  outside the stellar
extent of their host galaxy.  These lobes
contain hot spots near their outer edges.  The hot spots are identified with
the locations where the bulk velocities of the jets undergo rapid
deceleration in  shocks (Mach disks); the magnetic fields are compressed in these
shocks and the individual electrons (and perhaps positrons and/or protons)
are accelerated to very relativistic speeds at those locations, thereby
explaining the extraordinary emissivities (e.g., Blandford \& Rees 1974).  
Only very careful observations
allow the detection of a jet in FR II sources since they are often thousands of times
weaker than the lobes; very seldom is a second (counter-)jet found.
These FR II jets are usually very well collimated in comparison with
FR I jets (e.g., Jeyakumar \& Saikia 2000).

Readers interested in more thorough discussions of most of these
topics should first consult the older, but still very useful, reviews of 
Bridle \& Perley (1984) (primarily for the observations), and 
Begelman, Blandford \& Rees (1984) (primarily for the theory).
A superb summation of the entire subject can be 
found in the volume edited by Hughes (1991) and a substantial
recent review is by Ferrari (1998).  
Many excellent conference proceedings could be consulted, 
including those edited by Hardee et al.\ (1996), Biretta \& Leahy (2000) 
and   Laing \& Blundell (2001).
In light of the availability of these summaries, only a limited
number of classical contributions will be cited in this review, which
emphasizes some new work
selected from a vast literature. 

Radio emitting jets have also been identified emanating from a
few binary systems within our Galaxy; these have been dubbed
``microquasars''.  In addition, jets have been found emerging from
many protostars.  This brief review cannot
discuss these local manifestations of jets; the reader interested
in relativistic jets within our galaxy should
consult an excellent recent review
 by Mirabel \& Rodr\'iguez (1999).

\section{Some Key Observations and Implications}

\subsection{Properties of extragalactic jets}

The fact that synchrotron emission does not provide spectral lines
has meant that even the simplest facts about the properties of jets,
such as their velocities and compositions, have remained controversial.
While magnetic field structures can be directly probed through
polarization measurements, estimates of magnetic field strengths
can usually only be made through the assumption of equipartition
between the energies in the radiating relativistic particles and
the magnetic field energy density; this assumption has the advantage
of minimizing the total energy that needs to be supplied to the
radio source.

Evidence for relativistic bulk motion in the powerful FR II sources
has been generally accepted for quite a while, mainly because of the 
detection of apparent
superluminal expansions of pc-scale knots through Very Long
Baseline Interferometry (VLBI), along with the detection of 
fast motions on substantially larger spatial scales in a few sources.  
A source of radiation moving at
moving with velocity 
$\beta = v/c$ at an angle, $\theta$ to the
line-of-sight will have an apparent transverse velocity,
$\beta_{\rm app} = (\beta \sin\theta)/(1- \beta\cos\theta)$, which
can exceed 1 if $\beta \approx 1$ and $\theta$ is small.
Very fast motions, implying bulk Lorentz factors, $2 < \Gamma = (1 -
\beta^2)^{-1/2} < 10$, also naturally explain the substantial asymmetry
in the radio emissivities of extended jets in FR II sources; the
jet with a component towards us would be Doppler boosted,
while the (usually unobserved) counter-jet would be Doppler dimmed.
For a jet of intrinsic spectral index $\alpha$ [defined
so that $S_{\rm em}({\nu^{\prime}}) \propto (\nu^{\prime})^{-\alpha}$] the
observed flux would be (e.g., Scheuer \& Readhead
1979),
$S_{\rm obs}(\nu) = S_{\rm em}(\nu){\cal D}^{(2+\alpha)}$,  
where the Doppler factor, ${\cal D} = 
[{\Gamma}(1-\beta \cos
\theta)]^{-1}]$, and the observed frequency is related to the
emitted one by $\nu = \nu^{\prime}{\cal D}$; for small $\theta$,
this implies large enhancements.  For a single source, the exponent in the
above equation becomes $(3+\alpha)$, and the boosting effect is even more
pronounced.
Apparently abrupt changes of directions of jets on the VLBI scales are
also naturally explained in terms of small intrinsic direction changes
that appear magnified to distant observers by special relativistic effects.
Since the fluxes from the opposite hot-spots and lobes are usually very similar
in FR II sources, it is widely accepted that the bulk motions of
those extended regions are non-relativistic, with typical head
advance speeds of $\sim 0.03 c$ (e.g., Scheuer 1995).

Recent observations have found that many FR I jets also provide evidence of
 relativistic velocities 
on VLBI scales (e.g., Giovannini et al.\ 1994, 
Biretta et al.\ 1999, Xu et al.\ 2000).  
Doppler boosting can not only explain  that powerful FR~II jets appear one-sided,
but also that weaker FR~I jets exhibit large brightness asymmetry only
near their origins,  and typically
have short, one-sided basal regions (e.g., Laing et al.\ 1999; Kharb \& Shastri
2001). 
The evidence that FR~II jets retain relativistic bulk velocities out to
100's of kpc is quite convincing, with the brighter large scale jet always seen on
the same side as the nuclear jet and towards the less depolarized
radio lobe (e.g., Garrington et al.\ 1988, Bridle 1996, Gopal-Krishna
\& Wiita 2000a), so that
it is inferred to be approaching us.  However, for FR~I jets, their diffuse 
morphologies, the fact that the brightness asymmetry of the jets 
decreases with increasing distance from 
the core, and substantial large bends
frequently seen in FR I jets at multi-kpc scales imply that much slower flows
($v_{\rm bulk} < 0.1 c$) exist
on larger scales (O'Dea 1985, Feretti et al.\ 1999, 
Laing et al.\ 1999).

It is universally accepted that most, if not all, of the synchrotron emission
arises from relativistic electrons.  But the nature of the main positively
charged component of the jet plasma remains contentious.  
Until recently, the great majority
of workers have assumed that protons provided that neutralizing matter, and that
their much greater masses implied that they required a lot of extra energy to
accelerate but that they provided very little radio emission.  However, the
possibility that the jet was really composed of an $e^+$--$e^-$ plasma has
been suggested for a long time (e.g., Kundt \& Gopal-Krishna 1980).
Application of total energy and synchrotron radiation constraints led Celotti \& Fabian
(1993) to conclude that FR II jets were made of $e^-$--$p$ plasma,
since they argued that $e^-$--$e^+$ plasma of the required density
would yield too much annihilation radiation.  However,
Reynolds et al.\ (1996b) used similar energetic and radiation constraints
to conclude that the jet in the FR I source M87 was likely to be
made of $e^-$--$e^+$ plasma.  A similar argument favors an electron--positron
jet in the Optically Violently Variable Quasar 3C 279 (Hirotani et al.\ 1999).  If all of these arguments
are taken at face value, one might infer that the main difference between
FR I and FR II sources lies in the composition of the jet plasma,
and this would imply the existence of a fundamental difference
between their central engines.  But it is worth noting additional
 evidence for the presence of
pair plasma jets, even in FR II sources, comes from the interpretation of
the radio power--linear-size (P--D) diagram in terms of a model for 
quasi-self-similar growth 
of double radio sources (Kaiser et al.\ 1997).

Although there are  exceptions to this  rule, it is 
well established that, in general, the magnetic field 
in a FR~II jet remains aligned
with the jet along most of its length, while in a FR~I jet the magnetic
field is predominantly transverse on multi-kpc-scales
   (e.g.\ Bridle \& Perley 1984).  Careful studies of FR~I jets
have led to the conclusion that the asymmetries in apparent emission from 
the two jets, and their detailed
magnetic field patterns, are best explained if the
jets in these sources consist of
a narrow ``spine'' of relativistic flow
with a predominantly transverse magnetic field, surrounded by a slower
moving ``sheath'', probably contaminated by entrained material 
(a shear layer) 
where the magnetic field is stretched into a predominantly longitudinal 
configuration (e.g, Laing et al.\ 1999). 

\subsection{The Fanaroff--Riley Dichotomy}

There are many other observed differences between FR I and FR II radio
sources and the galaxies that host them (e.g., Baum et al.\ 1995,
Zirbel 1997, Gopal-Krishna \& Wiita 2000b).  All these observations
have led to the development of two general classes of explanations for the Fanaroff--Riley
dichotomy.

Intrinsic explanations involve a fundamental
difference in the central engine or jet properties 
between these two classes, while extrinsic explanations claim that
the differences arise through interactions of the jets with the media
through which they propagate.  Among the intrinsic explanations are
(see Gopal-Krishna \& Wiita 2000b for details and many more references):
the difference in jet composition mentioned above; a difference in the
central engine, such as having a more rapidly spinning black hole yield
FR II jets (e.g. Wilson \& Colbert 1995, Meier 1999); a difference in the accretion process, where advection dominated
flows might yield FR I jets, while more luminous ``standard'' accretion disks
might produce FR II jets (e.g. Reynolds et al.\ 1996a).  

The various extrinsic explanations
assume that the jets differ in little except total power or thrust.
In these scenarios, deceleration of the jet, through growth of
instabilities and/or entrainment of external plasma, converts weaker
jets into FR I morphologies, while stronger jets, which remain
supersonic and/or relativistic to great distances, produce FR II
structures (e.g., Bicknell 1984, 1995, Komissarov 1990). 
Recently Gopal-Krishna \& Wiita (2000b) have stressed that the existence
of a small number of sources with distinctly FR I morphologies on one
side of the core and FR II morphologies on the other side, can play an
extremely important role in distinguishing between these putative
explanations for the FR dichotomy.  They have found six good examples
of these HYbrid MOrphology Radio Sources, or HYMORS, and have argued
that while they are  expected to be rare if an extrinsic mechanism
dominates, HYMORS are unlikely to be found at all if an intrinsic
mechanism were to be important.

As more measurements were made of the galaxies that host radio jet
sources, it became clear that the simple radio source power criterion
found by Fanaroff \& Riley was not really appropriate.  Rather, as the
luminosity of the host galaxy, $L_{\rm opt}$ grows, so does the radio 
power, $P_R^*$, required to
produce an edge-brightened, FR II, morphology.  This was quantified
by Ledlow \& Owen (1996), whose extensive data compilations 
showed that $P_R^* \propto L_{\rm opt}^{1.7}$.
Bicknell (1995) demonstrated that this relation could be roughly reproduced within 
an extrinsic model for the FR dichotomy; in this picture, the weaker jets would
slow until $\Gamma \simeq 2$ and then suffer the growth of instabilities 
that lead to FR I type structures.  It has now been shown that a variant
of this scenario, using a somewhat different jet propagation 
 model (Gopal-Krishna et al.\ 1989), 
and where the trigger that yields FR~I structure is now the 
slowing of the advance speed of
the jet to subsonic with respect to the external ambient gas 
(Gopal-Krishna et al.\ 1996), can produce
an even better fit to the observed $P_R^*~-~L_{\rm opt}$ relation
(Gopal-Krishna \& Wiita 2001).  While both the magnetic-switch model
(Meier 1999) and the gravitational slingshot model (Valtonen \& 
Hein\"am\"aki 1999) can also yield rough agreements with the
radio--optical correlation, given the additional evidence from
the existence of HYMORS, it is clear that 
extrinsic explanations for the FR dichotomy are more likely to be correct.

\subsection{Implications of Relativistic Motions}

As mentioned in Section 2.1, the detection of apparently superluminal
transverse motions provides extremely strong evidence for
relativistic motions in jets, and these large $\Gamma$ values
also provide an excellent explanation of the preponderance
of asymmetric jet luminosities
in double radio sources with quite similar lobe powers.
The orientation at which we view these relativistic jets
can also provide an understanding of several other key features
of radio-loud AGN.

It is now widely accepted that AGNs with strong radio
jets will typically be classified as radio galaxies if the orientation
of the jet to our line-of-sight to the source is greater than
a critical value, $\theta_{\rm crit} \simeq 40^{\circ}$, while the same source will be called a
quasar if $\theta < \theta_{\rm crit}$  (e.g., Barthel 1989;
Urry \& Padovani 1995).  These unified models for radio-loud
AGN assume that the jets are launched parallel to the rotation
axis of a supermassive black hole (SMBH) and perpendicular to
the accretion flow feeding the SMBH.  Unified models also require the
presence of a thick dusty torus outside the accretion flow
(on the scale of several parsecs) that can absorb enough soft X-ray, UV and optical
radiation to hide both the direct core continuum emission and 
 the broad emission line region if viewed from angles above
$\theta_{\rm crit}$.  

In addition, if the jet is very close to our
line of sight ($\theta \simeq \Gamma^{-1}$) then very
substantial special relativistic effects would strongly enhance the
observed fluctuations and polarization.  Under these circumstances
 the source
might be classified as an Optically Violently Variable quasar
or  other type of blazar.  Extremely convincing explanations 
for the variations on the timescales of months of several quasars in 
terms of shock-in-jet models have been available for quite some time
(e.g., Marscher \& Gear 1985, Hughes, Aller \& Aller 1991), and very
rapid variability and polarization swings can be understood if
the shocks are travelling down a slightly bent jet (e.g.,
Gopal-Krishna \& Wiita 1992) or if there is strong turbulence
in the vicinity of the shock (e.g., Marscher \& Travis 1991).  
Recent measurements indicate
that much of the fastest (intraday) radio
variability (see Wagner \& Witzel  1995 and Wiita 1996 for reviews)
is probably due to interstellar scintillation (ISS; Rickett 1990);
two such sources are
 PKS 0405$-$385 (Kedziora-Chudczer et al.\ 1997) and
J1819$+$3845 (Dennett-Thorpe \& de Bruyn 2000).  Nonetheless,
extremely rapid intrinsic variations do appear to be required for
some sources.  One example is the BL Lac object 0716$+$714 where
there appears to be a correlation between variations in the
radio and optical bands (Wagner et al.\ 1996).  Another case is the 
gravitational lens system
B0218$+$357 (Biggs et al.\ 2001); here correlated variations between
the two images over
 a few days are seen to be nicely separated by the 10.5 day time lag
due to lensing, and cannot be explained in terms of ISS
or gravitational microlensing (Gopal-Krishna \& Subramanian 1991).

Turning to the weaker radio sources, there is now abundant evidence
that FR~I radio galaxies are the parent population for BL Lacertae
objects, in the sense that a typical FR~I source, if viewed at small
$\theta$, would show the properties of a blazar, and that the relative
numbers of these classes are nicely understood if this unification
holds (e.g., Urry \& Padovani 1995).  The synchrotron self-Compton
mechanism can neatly explain the overall spectral energy distribution
of blazars (e.g., Sambruna et al.\ 1996), but this only works if 
$\Gamma > 5$, for otherwise X-rays would be over produced.  Intrinsic
variability in many blazars implies  small linear sizes; these
translate into brightness temperatures
that substantially exceed the inverse Compton limit of
$\sim 10^{12}{\rm K}$ for incoherent synchrotron sources
unless similarly high values of $\Gamma$ are invoked.

As the sensitivity and dynamic range of radio observations have
continued to improve it has become clear that many so-called
radio quiet AGN are actually radio weak, but not radio silent, and that
there is a considerable population of radio-intermediate
sources (e.g., White et al.\ 2000).  
Quite a few Seyfert galaxies are now known to possess radio jets,
though they tend not to propagate very far, probably because of
more rapid disruption by the interstellar medium of their spiral
hosts (e.g., Pedlar et al.\ 1993) 
Again, a unified scheme seems to work very well, with the Seyfert
1 galaxies viewed at $\theta < \theta_{\rm crit}$ so the broad
line region and soft X-rays can be seen directly, while the
Seyfert 2 galaxies are viewed at $\theta > \theta_{\rm crit}$
and broad lines can only be detected in polarized reflected
light (cf.\ Antonucci 1993).

\section{Analytical Models for Jets}

\subsection{The launching of extragalactic jets}

While several implications from the observations of extragalactic
jets based on fundamental theories have already
been described, we now turn to matters less accessible to direct
observation because they occur on such minute spatial scales.  
Here we have space to merely list some of the ever growing lines 
of evidence
that have convinced essentially all astrophysicists that accreting
SMBH's provide the prime mover for AGNs: accretion into a very
deep potential well is the only mechanism that appears to be 
sufficiently efficient (easily $>$5\%) in converting matter into the 
energy to power the most
luminous AGN; fast luminosity variations imply most of the
emitted power must emanate from a very compact region, probably
just a few $r_g = GM/c^2 = 1.5~(M_{BH}/M_{\odot})$ km in extent; 
a SMBH, particularly a rotating one, is the only
known way to produce the stable axis needed to produce jets extending
for Mpc that must have remained active for $> 10^7$ years;
VLBI measurements have shown collimated emission to have been
produced within $< 100 r_g$ in a few nearby radio loud AGN; 
stellar velocity dispersions in the inner cores of nearby AGN rise
very steeply towards the center, implying immense densities of
matter ($> 10^7 M_{\odot}{\rm pc}^{-3}$); maser emission lines
have demonstrated clear Keplerian rotation about massive dark cores;
the shapes of x-ray emission lines are best explained as emerging
from the inner portions of accretion disks around SMBHs, where general relativistic
effects play important roles. 

We now proceed to a brief discussion of the various classes
of models that have been proposed to produce jets from the
environs of a SMBH.
The fundamental types of models proposed through the 1980's for the
origin of jets were reviewed extensively by Wiita (1991),
and here we will very briefly summarize those scenarios and
note some more recent results.
All models involve the expulsion of a certain fraction of
the matter being accreted by the SMBH.  The key distinction
between the major classes of jet launching scenarios is whether or 
not magnetic fields are assumed to be
primarily responsible for the expulsion of jet plasma.

Purely hydrodynamical models based upon winds from standard thin accretion disks 
(e.g., Shakura \& Sunyaev 1973)
have difficulties in accelerating substantial
amounts of matter to relativistic velocities.  Crudely, the maximum
velocity of an outflow depends upon the depth of the potential well
in the region from
which the matter is expelled, so if the matter is launched from a significant
portion of the disk, then much of it will have relatively low velocities.
Such a wind may be able to provide additional collimation to a jet
propelled from the innermost region of the AGN (e.g., Sol et al.\ 1989)
but is unlikely to provide either sufficient collimation or sufficient
velocity to explain the observations of VLBI scale extragalactic jets.

At very high accretion rates (comparable to or above that required to
produce the Eddington limit) radiation supported thick accretion disks
can form (e.g., Paczy\'nski \& Wiita 1980).  This type of
accretion flow provides narrow
funnels and a more centrally concentrated region from which to launch
the plasma, so it had promise to be a reasonable way to produce
powerful collimated beams.  This type of flow could produce
super-Eddington luminosities but involved quite low efficiencies,
as much of the emitted radiation was swallowed by the central
black hole in an extremely optically thick flow. However, once the matter 
gets to mildly relativistic speeds, the 
aberrated radiation actually produces a drag on the outflow and prevents
$\Gamma$ values much in excess of 2 (e.g., Narayan et al. 1983).
It might be possible to avoid this limit and attain $\Gamma \sim 10$
if clouds of electrons are present in the flow region; if they
can produce enough synchrotron self-absorption of the radiation from
the funnel, the red-shifted photons will be unable to decelerate the
flow (Ghisellini et al.\ 1990).  

Over the past decade other versions of low efficiency accretion flows,
which can occur for very low accretion rates,
have been frequently discussed.   Early proposals along these lines
(e.g.. Ichimaru 1977, Rees et al.\ 1982) have been developed along different
directions
 by Chakrabarti (e.g., 1990, 1996) 
and by Abramowicz,
Lasota, Narayan and collaborators (e.g., Narayan 
et al.\ 1998).  Aside from its fundamental importance as a way of
treating more general non-Keplerian accretion flows, the main rationale for the
of study of these advection dominated flows was
to explain the spectra of X-ray binaries in different
states and to fit the spectrum of radiation emerging from non-active
galactic nuclei, such as that in our Milky Way.  Recently, variants on these
models have been shown to accommodate substantial outflows (e.g.,
Begelman \& Blandford 1999).  When the velocity structure is
sub-Keplerian, as should happen 
under a wide range of 
accretion flows in the near vicinity of the BH, then the flow can quite naturally
lead to a centrifugal pressure supported
boundary layer, which has been shown to be capable of launching significant outflows 
(Das 1998, Das \& Chakrabarti 1999).
This mechanism appears to be able to produce adequately relativistic
outflows of reasonably good collimation.  However all of these 
fundamentally hydrodynamical (HD) models
make one or more critical simplifying assumptions, and their robustness and
range of applicability remain to be tested.

While the last mentioned HD launch mechanisms do
appear to be promising they have not yet been studied intensively, 
and at this point a large  majority of workers in
this area consider that magnetic fields have an important role to
play in the ejection and initial collimation of flows from the
vicinities of SMBH's.  The fact that jets emit via the synchrotron
mechanism makes it clear that magnetic fields are present, and several
plausible ways to use magnetic fields to accelerate and collimate
flows have long been known.  

The advantage of magnetic acceleration mechanisms is that they
can simultaneously and naturally produce relativistic velocities,
narrow jets and large momentum fluxes.
The idea that jets were
predominantly a Poynting flux with little mass loading was
proposed by Rees (1971), and the possibility that powerful currents
could be generated in accretion flows which would then
accelerate collimated outflows
was first noted by Lovelace (1976).  Other pioneering works in
this area were by Bisnovatyi-Kogan \& Ruzmaikin (1976) and Blandford
(1976).

An enormous number of variants on magnetically accelerated
jet models have been put forward over the past quarter-century,
and we cannot even begin to summarize this literature here.
But the majority of them fundamentally rely on either extracting
energy and angular momentum through magnetic fields anchored
in the disk (e.g., Blandford \& Payne 1982; BP), or by extracting
the spin energy of the black hole itself, through magnetic fields
threading its horizon (e.g., Blandford \& Znajek 1977; BZ).

An excellent introduction to the physics of  
magnetohydrodynamical (MHD) jet production mechanisms is the review by
 Spruit (1996).  The vast majority
of this research effort has naturally concentrated on ideal
MHD models (where the plasma is tied to the field lines) 
and makes the simplifying assumptions of stationary and
axisymmetry flows with infinite conductivity.  While the MHD assumption
does not always hold around pulsars, and the low temperatures
around protostars imply that finite conductivity can be important
there, for the conditions around BHs both of these assumptions
should be excellent.  Spruit (1996) shows that the centrifugal 
(beads-on-a-wire) approach
of, e.g.,  BP, and the purely
magnetic approach of, e.g., Lovelace et al.\ (1987)  are completely
equivalent.  

The self-consistent computation of MHD flows is extremely
difficult because of the possible presence of multiple critical points,
each of which can be associated with a shock, which can be of either
the slow- or fast-type (e.g., Heyvaerts \& Norman 1989).  
The types and locations of these
critical points depend sensitively on the assumptions made
about boundary conditions
and initial topology of the magnetic fields.  Nonetheless,
a variety of
initial conditions and analytical approaches have been explored
and do provide some general
conclusions.  MHD jets can collimate asymptotically to a cylindrical
structure if they carry a sufficient net current, whereas they are very likely
to attain a paraboloidal cross-section if they do not (e.g., Chiueh
et al.\ 1991).  Self-confined equilibria can be achieved and
analytically described in sensible 
 approximations (e.g., Appl \& Camenzind 1993). 

A recent generalization of the Blandford-Payne model allows for
a hotter initial plasma and finds solutions which start
 with a sub-slow magnetosonic speed and 
subsequently cross all critical points, at the slow 
magnetosonic, Alfv\'en and fast magnetosonic separatrix surfaces
(Vlahakis et al.\ 2000).  These models tend to over-collimate
toward the jet axis, as do many other MHD calculations, so it is
clear that some of the assumptions going into these models must be
relaxed.  Such relaxation can best be accomplished through the
numerical modeling to be discussed in Section 4.1.

It is worth recalling that extraction of the spin energy of the
SMBH through very low accretion rates coupled to magnetic fields 
is an alternative to unified models as a 
way of understanding the differences between
radio galaxies and quasars (Rees et al.\ 1982).  Recently
the underpinnings of the basic BZ mechanism have come under renewed
investigation.  Ghosh \& Abramowicz (1997) have argued that
magnetic field strengths in the inner parts of accretion disks
are weaker than estimated earlier so that the strength of the 
BZ process for the extraction of the rotational energy of
the black hole is lower than imagined previously.  Livio et al.\
(1998) argue that the magnetic fields threading the BH should
not be stronger than those in the inner parts of the disk, so that
the BZ mechanism should always contribute less power than the disk
feeding it.  However, if the field strength continues to grow in
the innermost disk region and if plasma plunging into the BH
exerts a strong torque on the innermost portion of the disk,
as has been suggested recently
(Krolik 1999, Agol \& Krolik 2000), then these caveats may be weakened.
Furthermore, it is worth noting that the standard BZ flow is likely to be subject to a
screw-instability which can limit the extent out to which it
can produce plasma acceleration (Li 2000a).   However, a related
scenario,  where magnetic 
field lines connect plasma particles inside the ergosphere of a Kerr BH
with remote loads, is of real interest.   Frame dragging 
twists the field
lines so that energy and angular momentum are extracted 
from the plasma particles, and if the magnetic field is
strong enough, then the particles can have negative energy
as they fall in, thereby allowing extraction of the
BH's rotational energy (Li 2000b).  In all of these efforts certain
crucial, and not yet adequately justified, assumptions must be
made; therefore, none of the specific MHD launch mechanisms can be considered
to be convincing, though many remain plausible.

\subsection{The propagation and stability of extragalactic jets}

Once launched, the key question becomes: can these theoretical jets
survive to the distances demanded by observations, where jet
hot-spots are often much narrower than 1\% of the jet length?
Stability analyses of hydrodynamical jets began with discussions
 of the Kelvin-Helmholtz (or
two-stream) instability which showed that faster jets,
particularly relativistic ones, could survive longer 
(e.g., Turland \& Scheuer 1976).  Important early contributions
were made by Hardee (e.g., 1982, 1987), Birkinshaw (1984), Ferrari et al.\
(1980), Bodo et al.\ (1989), among others.  Then the
analyses were expanded from two-dimensional cylinders to 
two-dimensional slabs, which can mimic some three-dimensional effects
(e.g., Hardee \& Norman 1988), to three-dimensional hydrodynamics
(HD); then, various assumptions about magnetic field geometry
have been studied within an ideal MHD framework.

The ordinary
mode ($N = 0$) is always excited whenever there is a boundary
between a flow and a static region.  However, various reflection
modes ($N > 0$), which exhibit $N$ pressure nodes within the
jet,  can also be excited if the walls of the cavity can vibrate
coherently; effectively, the sound waves hitting the boundary at
an angle can constructively
interfere within the jet. Under plausible circumstances the
$N = 1$ mode can grow faster than the $N=0$ mode.  Very typically,
the dominant modes in 2-D are those with wavelengths of $\sim 5 R_j$,
and these lead to growth lengths of $\sim 3{\cal M} R_j$,
where $R_j$ is the jet radius and ${\cal M}$ is the Mach number
of the jet (with respect to its internal sound speed).
The growth of both pinch ($m=0$) and kink ($m=1$) modes can be quite fast, 
but jets with smooth transverse velocity  gradients are more stable.

The stability properties of MHD jets have been recently addressed by many authors.
 Jet magnetic
field geometries that evolve into primarily concentric
toroidal structures are usually most unstable to kink ($m =1$) 
instabilities (Begelman 1998).  A  study of rotating jets
confined by toroidal fields has shown that rigid rotation tends
to stabilize, while differential rotational destabilizes, the 
jet in a way similar to the magneto-rotational instability which
is now believed to dominate viscosity production in accretion disks
(Hanasz et al.\ 2000).  In this local analysis, if the azimuthal
velocity exceeds the Alfv\'en azimuthal speed, the rigidly rotating 
part of the jet interior can be completely stabilized, while the strong
shearing instability acts on the  layer between the 
rotating jet interior and the external medium, perhaps thereby
explaining the limb-brightening seen in some jets (Hanasz et al.\ 2000).
Other rotating MHD jet models have been recently analyzed for
stability by Lery \& Frank (2000). These connect to a Keplerian disk
and have a complex structure: a dense, current carrying central core; an 
intermediate magnetically dominated region; and a low density outer region
carrying a return current.  Another approach to the stability of
rotating MHD jets has been taken by Kersal{\'e} et al.\ (2000), who
use the ballooning ordering expansion to find that cylindrical
configurations can be destabilized by a negative magnetic shear as well 
as by a favorable equilibrium pressure gradient.  They note that
rotating jets with vanishing current density along the axis, as well as most
non-rotating MHD jet models, would be unstable.

The major shortcoming of all of these analytical models is that
they can only compute the linear growth rates of various instabilities
under initially regular conditions; if taken at face value,
probably no analytically computed
jet could propagate stably for 100 kpc or more, yet of course many such
extended extragalactic sources are observed.  
Therefore high-resolution numerical simulations
are required to explore the non-linear effects which can provide 
saturation of the linear instabilities.

\section{Numerical Simulations}

A good summary of many results emerging from HD and MHD
computations of jets was given in Ferrari (1998). In this 
section we concentrate on more recent simulations. 

\subsection{The launching of extragalactic jets}

Early efforts to incorporate GR effects in simulations
of accretion onto a BH indicated the likelihood of significant
outflow, even in purely HD situations (Hawley et al.\ 1984).
However, until very recently, this work and other attempts along the
same lines were
greatly hampered by severe numerical difficulties.  These demanded
the development of techniques, such as adaptive mesh refinement,
that allow one to efficiently and simultaneously compute flows in the
high density regions in the
accreting gas and in the low density expelled gas; following the
latter also 
requires much greater spatial scales which make the computations
extremely expensive.

Pioneering work on MHD launching of jets from disks was
performed by Uchida \& Shibata (1985), 
who evolved an initially
vertical magnetic field tied to a cold
thin disk rotating around a point mass assuming axisymmetry.  
Differential rotation in the disk
produces a substantially toroidal field and this magnetic tension
is released through strong torsional Alfv{\'e}n waves, which expel mass.
This approach has recently 
been extended to 3-D relativistic flows around Schwarzschild BH's
by Koide et al.\ (1998) and
Nishikawa et al.\ (1999). 
They find that a shock forms in the disk and yields a gas-pressure
driven jet which dominates the outflow, though a weaker MHD
jet is present outside the pressure driven jet.
In a truly impressive computation, this work has recently been
extended to the environs of a rapidly rotating Kerr BH (Koide et al.\ 2000);
while the results for a corotating disk do not greatly differ from
those of the Schwarzschild situation, for (the relatively unlikely case of)
counter-rotating disks a very powerful magnetically driven jet is 
formed inside the gas-pressure driven jet.

The launching of cold gas from a disk under circumstances carefully designed to
emulate the BP magneto-centrifugal mechanism has recently been simulated
in 3-D by Krasnopolsky et al. (1999).  If the field is set up to be
``propelling'' then rapid acceleration and collimation of the flow
are indeed observed.  A simulation of the situation where a Keplerian
disk is initially threaded by a dipolar poloidal magnetic field has been
recently performed by Ustyugova et al.\ (2000); they find that a quasi-stationary
collimated Poynting jet arises from the inner part of the disk, while 
a steady uncollimated hydromagnetic outflow emerges from 
the outer part of the disk.  Although these calculations are
focussed on the types of overdense cooling jets that are to be found
in protostellar systems instead of AGN, it is also worth noting the
sophisticated numerical techniques involved in the simulations of 
Stone \& Hardee (2000).

\subsection{The propagation and stability of extragalactic jets}

Early 2-D simulations of HD jets (e.g., Norman et al.\ 1982)
were of great importance in establishing
that extragalactic jets were of very low density and of high Mach number,
for the morphology of FR~II radio galaxies could only be
reproduced under those circumstances.  The jet is preceded by
a bow shock; the cocoon is comprised of shocked ambient medium,
separated by a contact discontinuity from jet material that has
passed through a Mach disk shock at the head of the jet, which
corresponds to the hot-spot.  Since then, as the largest
computers have been turned to this task, the computations have
greatly improved in both spatial resolution and temporal duration.
Very long term 2-D simulations, which allowed the growth of
axisymmetric Kelvin-Helmholtz instabilities to go non-linear
(typically after the jets propagated distances corresponding to
hundreds of initial radii) indicated that
the lobes could become detached from the jets, but that new
Mach disks could form behind them, thereby explaining some of
the ``double--double'' radio source morphologies (Hooda et al.\ 1994).
A suite of 2-D relativistic and nonrelativistic jets have
recently been compared to show that the velocity field of nonrelativistic 
jet simulations cannot be scaled up to give the spatial 
distribution of Lorentz factors seen in relativistic simulations,
as had been often speculated to be the case 
(Rosen et al.\ 1999); however, each relativistic jet and its nonrelativistic
equivalent do have similar ages, if expressed in the appropriate dynamical
time units.

Three-dimensional simulations have clearly shown that non-axisymmetric
instabilities will become important if even small perturbations are
applied (e.g., Hooda \& Wiita 1998).  Nonetheless, the HD
jets can propagate to very substantial distances without completely
breaking up if they have high enough Mach numbers.  A careful
comparison of numerical simulations and normal mode analysis for
relativistic 3-D jets has shown that a wide variety of helical modes
can be generated; these imply that dramatic variations in Doppler boosting
are possible without much overall bending of the jet (Hardee 2000).  
Higher resolution simulations of
relativistic jets indicate that the instabilities are greatly
reduced in comparison to nonrelativistic situations (Aloy et al.\ 1999).
Other relativistic simulations have convincingly 
shown that the knot structures seen
in VLBI observations can be reasonably reproduced in terms of shocks
within those jets (e.g., Mart{\'i} et al.\ 1995, Mioduszewksi et al.\
1997, G{\'o}mez et al.\ 1998). 

 The collision
of a jet with a much denser cloud have recently been reexamined using
high resolution 3-D HD simulations
(e.g., Higgins et al.\ 1999, Wang et al.\ 2000).
While powerful jets will destroy most obstructions and weak jets will
be stalled and destabilized by them (as probably happens in many Seyfert
galaxies), there is a rather small region
of parameter space where jets can bend and survive; this could
explain some rare ``dog-leg'' morphologies.

The instability of MHD jets, particularly focussed on the 
question of entrainment, has been carefully studied under
various situations recently (Hardee \& Rosen 1999, Rosen \& Hardee 2000). By precessing the jets at the origin to excite the KH
instability, results can be compared with linear stability analyses,
and it is concluded that the KH instability is the primary cause 
for mass entrainment
but that expansion of the jet reduces the rate of mass entrainment. 

An interesting approach to MHD jet stability has been taken by Frank et al.\
(2000).  The initial conditions for the jets are taken from analytical
models for magneto-centrifugal launching and have a more complicated
structure than most earlier work.  They find new behavior 
including the separation of an inner jet
core from a low density collar.  The wavelengths and growth rates
from a linear stability analysis are in good accord with
2.5 dimensional numerical simulations (Lery \& Frank 2000).
For a sub-class of current-driven instabilities in cold
supermagnetosonic jets, 3-D MHD simulations
have also found good agreement with a linear analysis (Lery et al.\ 2000).
If the initial equilibrium structure has a pitch profile that 
increases with radius, an
internal helical ribbon with a high current density forms,
which yields localized dissipation; this might produce
particle acceleration within the jet.

\section{Conclusions}

Jets are ubiquitous.  As astronomical instrumentation has improved
we have been able to detect jets over an absolutely phenomenal range
of distances and powers.  Extremely rapid flows in extragalactic
sources explain extraordinarily fast variability, 
apparent superluminal motions, and the spectra
of blazars over 17 decades in frequency.  Our viewing of these
jets from different angles can explain essentially all
of the differences between FR II radio galaxies and radio-loud
QSOs as well as the differences between FR I radio galaxies
and blazars.  

No specific model for the production of cosmic jets is absolutely
compelling.  While it is highly likely that MHD processes are
of importance in the launching and initial collimation of jets,
the details of these processes remain extremely controversial. 
Given the complexity of MHD in full general relativity, this
is not surprising.
Advances in numerical techniques and computing power are finally
 allowing
tentative explorations of 3-D relativistic MHD flows.  
However, only when several
groups, using different codes and wide ranges of plausible initial
conditions, all produce very similar outcomes,  will it
be fair to claim that the source of jets in AGN is understood.

More confidence can be given to the results on propagating jets,
since the results of simulations can be reasonably matched to
observations.  The idea that deceleration through interaction with
the external medium converts FR~II type jets into FR~I types appears
to be valid.  Nearly all modern 3-D simulations, whether HD or MHD, 
whether relativistic or not, tend to produce flows that can 
propagate over
very long distances, though instabilities can yield a ``flailing
about'' of the outer portions of the jet.  The interaction of the jet with the external 
medium always provides a sheath of matter that can assist in confining
the jet and appears to be able to slow the growth of unstable modes.

Although we have not had the space to discuss them, we must end by
 recalling
that much smaller and weaker,
but still relativistic,
outflows that can be legitimately characterized as jets are found
in some compact binary systems (Mirabel \& Rodr{\'i}guez 1999).  Furthermore, reasonably collimated, 
albeit much
slower, flows are common around young stellar objects (YSOs; 
for a recent review see
Richer et al.\ 2000).  The argument
that all of these collimated outflows 
are formed in very similar ways (e.g., Livio
1999) is intriguing, but by no means convincing.  The differences
in physical conditions are so immense, and our understanding
of the origin of extragalactic jets so tenuous, that 
any claims along these lines are most speculative.  It is at least
as likely that a substantially different physical mechanism, such
as the X-wind model (Shu et al.\ 2000)
dominates the slow, weak jets in YSOs and differentiates them from
cosmic radio jets.

This work was supported in part by NASA grant NAG 5-3098, by
Research Proposal Enhancement funds at Georgia State University
and by the Council on Science and Technology at Princeton University.

\end{document}